\documentclass[letterpaper, 10 pt, conference]{ieeeconf}  
\usepackage{graphicx}
\usepackage{hyperref}

\IEEEoverridecommandlockouts                              

\title{\LARGE \bf
Realtime Estimation of IEEE 802.11p for Mobile working Machines Communication respecting Delay and Packet Loss\\
}

\author{Yusheng Xiang$^{1}$, Tianqing Su$^{2}$, Xiaole Liu$^{3}$ and Marcus Geimer$^{1}$
\thanks{$^{1}$Yusheng Xiang and Marcus Geimer are with Institute of Vehicle System Technology,
        Karlsruhe Institute of Technology, 76131 Karlsruhe, Germany
        {\tt\small yusheng.xiang@partner.kit.edu}}%
\thanks{$^{2}$Tianqing Su is with the Institute of Communication Technology, Technical University of Braunschweig,
        38106 Braunschweig, Germany
        {\tt\small t.su@tubs.de}}%
\thanks{$^{3}$Xiaole Liu is with the Department of Electrical and Computer Engineering, Technical University of Munich,
        80333 Munich, Germany
        {\tt\small xiaole.liu@tum.de}}%
}

\begin{document}

\maketitle
\thispagestyle{empty}
\pagestyle{empty}

\begin{abstract}

The fleet management of mobile working machines with the help of connectivity can increase not only safety but also productivity. However, rare mobile working machines have taken advantage of V2X. Moreover, no one published the simulation results that are suitable for evaluating the performance of the ad-hoc network at a working site on the highway where is congested, with low mobility, and without building. In this paper, we suggested that IEEE 802.11p should be implemented for fleet management, at least for the first version. Furthermore, we proposed an analytical model for machines to estimate the ad-hoc network performance, i.e., the delay and the packet loss probability in real-time based on the simulation results we made in $ns-3$. The model of this paper can be further used for determining when shall ad-hoc or cellular network be used in the corresponding scenarios.

\end{abstract}

\section{Introduction}

Besides artificial intelligence \cite{c0}, the fleet management of mobile machines is the principal research direction of the internet of things in the fields of mobile working machinery.  Currently, the mobile machines are distributed sparsely in the working site and working at low transport speed to avoid a collision. With the vehicle-to-everything (V2X), the information about current position, speed or even destination and task are exchanged periodically between each individual mobile machine. Since the intentions of neighbor mobile machines within sensing range are known, the working machine can work more densely and transport the material more efficiently. The most challenging and research-worthy use case can be described as the task of repairing the highway. During repairing the highway, a traffic congestion is usually expected. According to the study from Triantis, traffic congestion causes significant economic losses \cite{c1}. Apparently, by investing more machines with the help of V2X technology in a particular site can surely improve the working productivity, so that the economy lost due to the congestion can be diminished. Assuming that, all or part of the vehicles are equipped with V2X, a high channel load of V2X network occurs in the traffic congestion. Thus, the V2X performance decreases, manifesting in larger delay and packet loss probability.  In this paper, we first evaluate the performance of the IEEE 802.11p standard for varying node density rates by means of simulations using $ns-3$ \cite{c2}. Since the simulation model is computationally expensive, we then propose an fast estimation model for mobile machines to predict the mean delay and package loss probability of the IEEE 802.11p-based V2X network.

Fig.\ref{fig:Comparison the working site with/without V2X: more mobile working machines in the site, much higher productivity} illustrates the benefits of the implementation of V2X technology on mobile machines.

   \begin{figure}[thpb]
      \centering
      \includegraphics[width=3.2in]{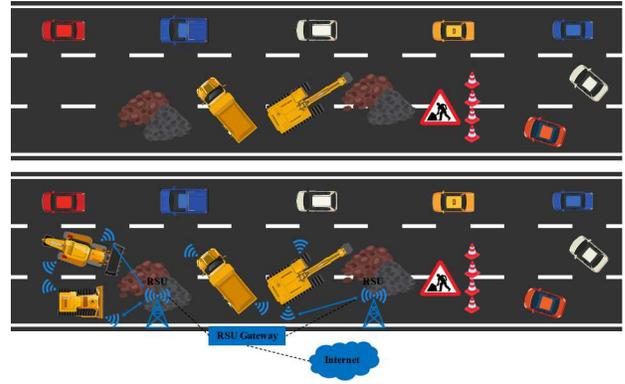}
      \caption{Comparison the working site with/without V2X: more mobile working machines in the site, much higher productivity}
      \label{fig:Comparison the working site with/without V2X: more mobile working machines in the site, much higher productivity}
   \end{figure}

\section{Current Wireless Communication for V2X}

The time-efficient and reliable message exchange among vehicles have been a longstanding issue for Intelligent Transportation System (ITS), which aims at enhancing the driving safety management as well as fulfilling requirement for infotainment service. Currently, there are two common used technologies for V2X, IEEE 802.11p and 3GPP Cellular-V2X \cite{c3}. IEEE 802.11p is the first standard for vehicular communication \cite{c4}. Both ITS-G5 and the Wireless Access in Vehicular Environments (WAVE), which is proposed by the EU and the US separately, amend the IEEE 802.11 standard for vehicular use \cite{c5}.

In the last two decades, the tremendous evolution of wireless communication technique has paved the way for the materialization of ITS. In 1999, 75 MHz of free but licensed spectrum at 5.850-5.925 GHz was allocated by US Federal Communications Commission (FCC) for implementation of the Dedicated Short Range Communications (DSRC) exclusively for the vehicle to vehicle/infrastructure communications. In the US, the spectrum is divided into seven 10 MHz channels with 6 Service Channels (SCHs) and a Control Channel (CCH). Compared with the US, the European Union (EU) introduced five channels (5.875-5.925 GHz), where CCH is restricted to safety usage only \cite{c6}, i.e., Cooperative Awareness Message (CAM). CAM is a periodic broadcast message which contains safety-relevant information, such as position, speed, acceleration. Until the time when the author writes this paper, the final version of the IEEE 802.11p is the version published in 2010 \cite{c5}. IEEE 802.11p is an ad-hoc network that has a mesh topology and thus has shortages such as a limitation to the short communication range, the medium mobility, as well as the contention. The coverage of IEEE 802.11p mainly depends on the transmit power \cite{c7}, path loss, signal fading, delay spread, Doppler spread, and angular spread \cite{c8}. The delay is unbounded, caused by carrier sense multiple access with collision avoidance (CSMA/CA) \cite{c9}.

In comparison with the WLAN-based IEEE 802.11p,  C-V2X uses the cellular networks and thus the communication relies on base stations. C-V2X uses 3GPP standardized 4G LTE or 5G mobile cellular connectivity \cite{c10}. As Vukadinovic pointed out, the C-V2X is a developing technology, from 3G to 5G \cite{c11}. With a supervised star topology, the collision of information is avoided. However, an obvious shortage of cellular network is the relative high delay even under a low channel load due to the round-trip between transceiver nodes and the base station. In release 14, 3GPP introduced direct vehicle-to-vehicle (V2V) communication outside of coverage under LTE-V mode 4 \cite{c12}. However, the distributed scheduling for LTE-V mode 4 is principally cannot totally avoid collisions. As the best of author's known, a congestion avoidance mechanism from 3GPP doesn't outperform IEEE 802.11p.

\section{Why We Use the IEEE 802.11p?}

Despite the fact that LTE has a series of advantages, we would like to adopt the IEEE 802.11p as our first version for connected mobile machines due to the following reasons. First of all, to fully make the advantages of C-V2X, mobile machines need a base station nearby, which varies from 10m until 10km \cite{c13}. However, for the fleet of mobile machines that are working far away from urban, they might fail to find a base station nearby. Moreover, the usage of 802.11p is free of charge. Different from the cellular network which the users must pay for the service from the network operators, the 5.9 GHz band is a free but licensed spectrum \cite{c4}. In addition, IEEE 802.11p is well designed for the vehicle industry so that no additional modification is needed for vehicle onboard ECU \cite{c14}. Thus, the compatibility of IEEE 802.11p is better for the mobile machine which is designed without the consideration of V2X. Usually, mobile machines drive at a relatively lower speed. Furthermore, the communication between other on-board units, for instance, driving cars and mobile machines is not essential; thus, the under-performed ability to deal with vehicle mobility by IEEE 802.11p, based on the analysis of Alasmary's study \cite{c15}, can be ignored.
Although there have no consensus about which wireless technology is the more promising technology, scientists from both sides agree that the combination of LTE and 802.11p have a certain improvement in performance compared to if only one technology is used \cite{c3,c7,c14,c16}. Thus, we would like to use IEEE 802.11p as the communication technology for our initial version fleet management. Even though the passenger car industry adopts cellular technology in the future, the idea of using IEEE 802.11p for mobile machines is still sensible, because the congestion of the channel is consequently alleviated.

\section{Modelling}

Mecklenbr{\"a}uker has shown the common scenarios in their paper \cite{c8}. Unfortunately, for mobile machines that have the task to repair the highway, the scenario does not belong to these common ones. Firstly, there has usually no buildings around the working site, but the traffic is congested. Secondly, instead evaluate the communication among all the participants in the ad-hoc network, only communication among mobile machines is essential. 

\subsection{Propagation Model}

In \cite{c17}, a comparative analysis between different propagation models is performed. Based on Stoffer's study, there is no best model for all cases, and the users should select the model depending on the concrete use case. Because we are mainly interested in delay and packet loss result congestion control algorithms at MAC layer and the highway is more similar to an urban scenario, we used a log-distance path loss model proposed by \cite{c18}. It is denoted as 

\begin{equation}
PL(dB)= PL(d_0) + 10nlog(\frac{d}{d_0})\label{eq}
\end{equation}
where PL(d0) is defined as the path loss at the reference distance (d0), and $PL(d0) = 46.6777$dB. n refers to the path loss distance exponent varying from the propagation environment, and $n = 3$.

Since the single factor that influences receive power is the distance from the transmitter, in the following simulations, the dynamic mobility model is not applied to vehicles. Still, the relative positions of the vehicles are randomly initialized. 

\subsection{CAM’s Generation Model}

Venel presented that CAMs are generated at a rate in a range of 2 to 20 packets /second corresponding to multiple factors such as driver’s reaction time and vehicle speed \cite{c19}. Thereby, we apply a mean value from them, namely 10 packets/ second (10 Hz). In addition, the length of a packet varies from different applications in real-world vehicular communications. In the following simulations, packet length is set to be 450 bytes, which ensures the necessary information for the safety-related application. Since the generation rate and CAM length are constant throughout the simulation, the channel load is only depended on the number of nodes in the scenario.

\subsection{CSMA/CA and  Enhanced DCF Channel Access (EDCA)}\label{AA}

CSMA/CA algorithm is specified in IEEE 802.11 to schedule transmissions over a single channel by differing the access attempt with a random back-off time. In the meantime, EDCA introduces Interframe Spaces (IFS) and different contention window size to prioritize access categories and to improve quality-of-service (QoS) \cite{c20}.

Since the primary emphasis of this paper is on the congestion control algorithms at MAC layer and CAM length is constant, the term delay in the following part will always refer to the back-off time between the time point that a node request for channel access and the packet is forwarded from the MAC layer to the PHY layer, neglecting the transmission time depending on packet length and propagation time depending on distance.

Table \ref{ft_tab_ex} contains the vital parameters setting that we use.

\begin{table}[!h]
	\centering
	\caption{Simulation parameters}
	\begin{tabular}{l|cl}
	\hline \hline
	Parameters
		& Value
		& Unit\\ \hline
	TxPower
		& 17 
		& dBm \\
	Packet length
		& 450
		& Bytes \\
   	Packet generation rate
		& 10
		& Hz \\
	Channel width
		& 10
		& MHz \\
	Data rate (BPSK)
		& 3
		& Mbps \\
	Data rate (QPSK)
		& 6
		& Mbps \\
	CWmin
		& 15
		& - \\
	AIFSN
		& 7
		& - \\
   Time Slot
		& 13
		& $\mu$s \\	
	SIFS
		& 32
		& $\mu$s \\	
	EIFS
		& 120
		& $\mu$s \\

	\hline \hline 
	\end{tabular}
	\label{ft_tab_ex}
\end{table}

There are two ranges, i.e. transmission range and sensing range for each transmitter, since the CAM header and payload is modulated with different schemes and different immunities against noise and channel fading. The Physical Layer Convergence Protocol (PLCP) header is modulated with Binary Phase Shift Keying (BPSK) \cite{c21} and the payload is transmitted in the form of Quadrature Phase Shift Keying (QPSK) modulation,  Simulation results show that, the transmission range is equal to 115 meters corresponding to a SINR level at 6.49825 dB and the sensing range is equal to 175m. Once two transmitters are distanced more than 175m, they can send packets simultaneously, being unconscious of the busy channel status. In this case, they are called Hidden Node.  Multiple arbitrary packets may collide at the receivers who are visible and connectable to both hidden nodes. The interference between each other results in transmission failures.

In short, the scenario we analyzed is a working site on the highway where the communication performance among mobile machines under the interference from cars nearby.

\section{Evaluation of Hidden Node Problem}

To evaluate the impact of hidden node problem on vehicle network, a set of simulations is considered as follows: A total of 80 neighbor nodes is equally divided into two groups, which are symmetrically distributed on both sides of Transmitter/Receiver pair (Tx/Rx). 12 simulations are executed, with the distance between two groups of neighbors increases by 20 meters from 0 to 220 meters and 300 CAMs are sent per each node.
   
\begin{figure}[thpb]
      \centering
      \includegraphics[width=3.5in]{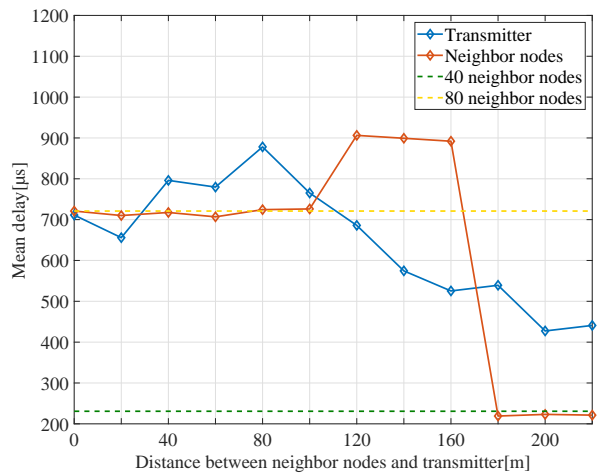}
      \caption{Mean delay[$\mu$s] versus distance between two groups of neighbor nodes[m]}
      \label{fig:MeanDelayHiddenNodes}
   \end{figure}

\begin{figure}[thpb]
      \centering
      \includegraphics[width=3.5in]{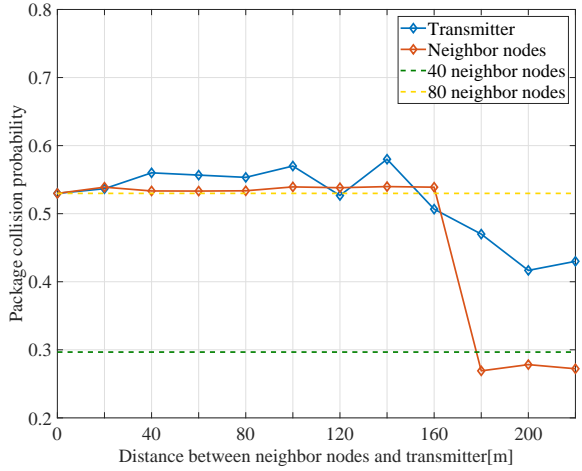}
      \caption{Packet collision probability versus distance between two groups of neighbor nodes[m]}
      \label{fig:CollisionProbaHiddenNodes}
   \end{figure}

\begin{figure}[thpb]
      \centering
      \includegraphics[width=3.5in]{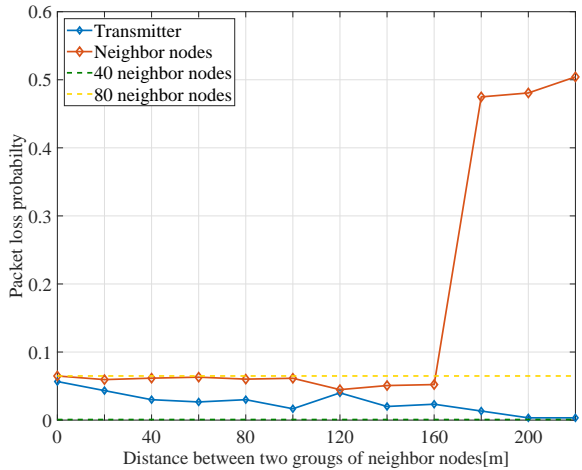}
      \caption{Packet loss probability versus distance between two groups of neighbor nodes[m]}
      \label{fig:LossProbaHiddenNodes}
   \end{figure}

How the different distance of two neighbor groups impacts the mean delay, packet collision probability, and packet loss probability of transmitter and neighbor are demonstrated in Fig. \ref{fig:MeanDelayHiddenNodes}, Fig. \ref{fig:CollisionProbaHiddenNodes}, and Fig. \ref{fig:LossProbaHiddenNodes}, individually. Performance observed at the transmitter and neighbors are illustrated with blue and red curve, respectively. As reference, the yellow and green dotted lines indicate the simulation results in which 40 and 80 neighbors are located at the same position as the transmitter. 

With respect to mean delay in Fig. \ref{fig:MeanDelayHiddenNodes},  the curve for neighbors remains stable within 115m and then rises in the sensing range owing to the additional EIFS appended to AIFS. Finally, it sinks significantly when the two groups are more than 175m apart from each other. In this case, they are hidden to each other. Therefore, the delay in each group is approximate to the scenario with just 40 neighbor nodes in the transmission range. In the meanwhile, the curve for transmitter fluctuates slightly. The reason is that the mean delay of transmitter is averaged by 300 packet in contrast with $80\times300$ packets of neighbors.The mean delay of the transmitter decreases when two neighbor groups are in each others' sensing range because the higher delay of the neighbors provides the transmitter a higher probability to access the channel. When the neighbors are hidden to each other, transmissions from hidden nodes overlap with each other, the whole channel busy time decreases. As a result, mean delay of transmitter declines. 

Similarly, the packet collision probability, which solely depends on the number of sensible nodes, are shown in Fig. \ref{fig:CollisionProbaHiddenNodes}. The red curve for neighbors remains coincident with 80 neighbors’ scenario and grows down rapidly to the 40 neighbors’ level as the two groups become hidden nodes to each other. In the meanwhile, the collision probability of the transmitter keep steady until the neighbours become hidden nodes. Since more idle channel is released due to overlapped transmissions, as mentioned in the previous paragraph, the packet collision probability of the transmitter declines, as wells as its packet loss probability, which is shown in Fig. \ref{fig:LossProbaHiddenNodes}.

The overlapped transmissions from hidden nodes packets are collided and corrupt at the receiver, resulting in a dramatic growth on packet loss probability of the neighbor nodes, which can be clearly seen from the red curve in Fig. \ref{fig:LossProbaHiddenNodes}. In the meanwhile, the transmitter has less collided transmissions. In brief,  the transmitters benefits from the appearance of neighbor nodes in form of hidden nodes in pairs, in terms of a less mean delay, packet collision probability, and packet loss probability.

The factor number of neighbors has a significant impact on the network performance, particularly in the case that packet length and generating rate is fixed.

\section{Empirical Model for Fast Estimation of Ad-hoc Network Performance}

Although $ns-3$ can simulate the V2X performance regarding the delay and the probability of lost packet, we still need a quick estimation method, so that on-board ECU can obtain V2X performance in real-time and evaluate the plausibility of V2X data. Therefore, we build an empirical model to fast estimate the network performance based on the results from $ns-3$. Since the contention behavior due to CSMA/CA in corresponding ranges should follow the same roles, which highly depend on the number of neighbors, we introduce the analytical model as follows.

\subsection{LuT Generation}
For each Cluster, e.g., the area within the transmission range and the area between the transmission and sensing range, we generate a Lookup-Table (LuT) in advance, which contains a set of crucial performance indicators in relationship with varying number of neighbors. To reduce the effect of randomness, we average the results from a large number of CAM transmissions.

To generate LuT for 1 cluster, we execute the following simulations. The neighbors are located at the same position with 60 meters away from the transmitter. The number of neighbors varies from 5 to 200, with a step of 5 in each scenario. Furthermore, for each of the 40 scenarios, 5 simulations are conducted, in which every single node schedules 1000 transmissions. The same simulations are executed for the 2. LuT, only the neighbors are 140m away from the transmitter.

   \begin{figure}[thpb]
      \centering
      \includegraphics[width=3.5in]{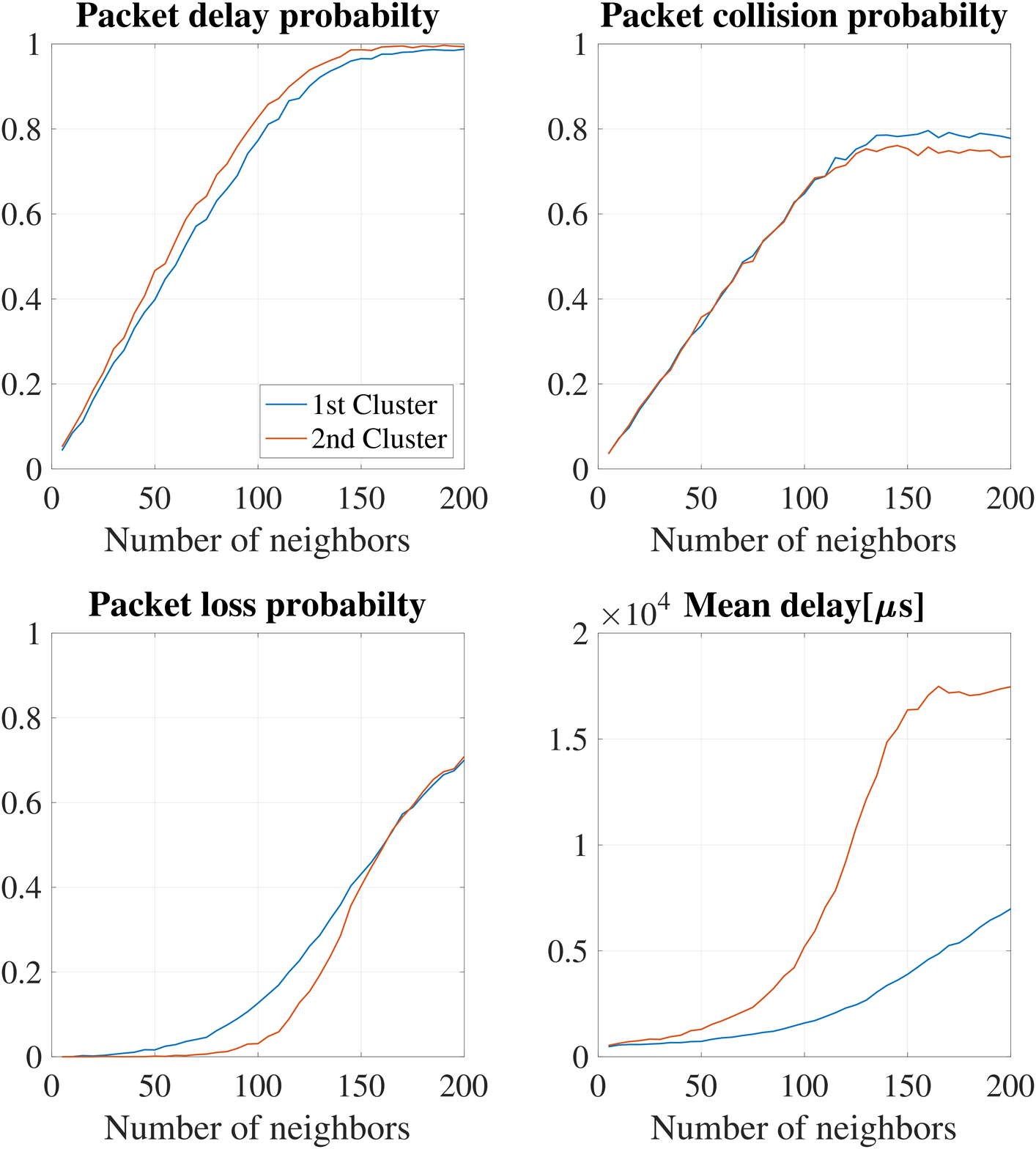}
      \caption{Packet delay probability, packet collision probability, packet loss probability and mean delay measured with varying number of neighbors in 2 clusters are included in the LuT}
      \label{fig:LUT_4}
   \end{figure}

Four metrics of the transmitter are measured, as shown in Fig. \ref{fig:LUT_4}, e.g., collisions probability ($P_{c}$), packet delay probability ($P_{d}$), packet loss probability ($P_{l}$), and mean delay ($t_{md}$). The term collision indicates the access attempt occurs during the duration, in which another node is transmitting. Moreover, the access attempt can also be differed due to the on-going AIFS, which follows the previous transmission, even though the channel is idle. Therefore, the percentage of delayed packets is slightly higher than the percentage of collisions. The metrics packet delay probability and mean delay indicate how probable the packet would be delayed due to an access contention, and once delay occurs, what would be the average duration.

\subsection{Performance Estimation}

For each on broad unit in the scenario, the number of neighbors located in each of the two Clusters are measured. The analytical result is derived from the sum of two values that are interpolated and extracted from LuTs. Furthermore, the upper limit for an analytical percentage is equal to 1. Eq. \ref{eq:NaiveEstimation,t} and Eq. \ref{eq:NaiveEstimation,p} demonstrates this idea,
  
\begin{equation}
\hat{{\Phi}}_{A,t} = LoU_{t,1}(n_T) + LoU_{t,2}(n_S)\label{eq:NaiveEstimation,t}
\end{equation} 

\begin{equation}
\hat{\tilde{\Phi}}_{A,p} = min(1, LoU_{p,1}(n_T) + LoU_{p,2}(n_S)\label{eq:NaiveEstimation,p}
\end{equation}
where $\hat{\tilde{\Phi}}_A $ is the naive estimation of the performance of the ad hoc using the analytical model, the footnote $t$ and $p$ denote the estimation in terms of time and probability, respectively.  $n_T$ is the node numbers inside of transmission range, $n_S$ is the node numbers inside of sensing range.

\section{Validation and Calibration}

In this section, we first validate the viability of the analytical model and then introduce the correction factor to eliminate the error between the naive LuT and the realistic simulation results. 

In the validation simulation, the traffic scenario is set to be a 1500m long highway with 3 lanes in each direction. 500 on-board units equipped with 802.11p devices are located statically. Congested traffic due to a highway worksite is assumed. The simulation is set up with a total simulation time of 100s, in which the vehicles are randomly distributed on the road.

The delay relevant metrics are simulated and estimated among all on-board units. This is because each transmission has a unique channel access time, which is independent of reception. In the meanwhile, for each on-board unit, the packet loss probability is measured on a single receiver, which is located randomly within its 15m range, corresponding to two cooperating mobile machines.

   \begin{figure}[thpb]
      \centering
      \includegraphics[width=3.5in]{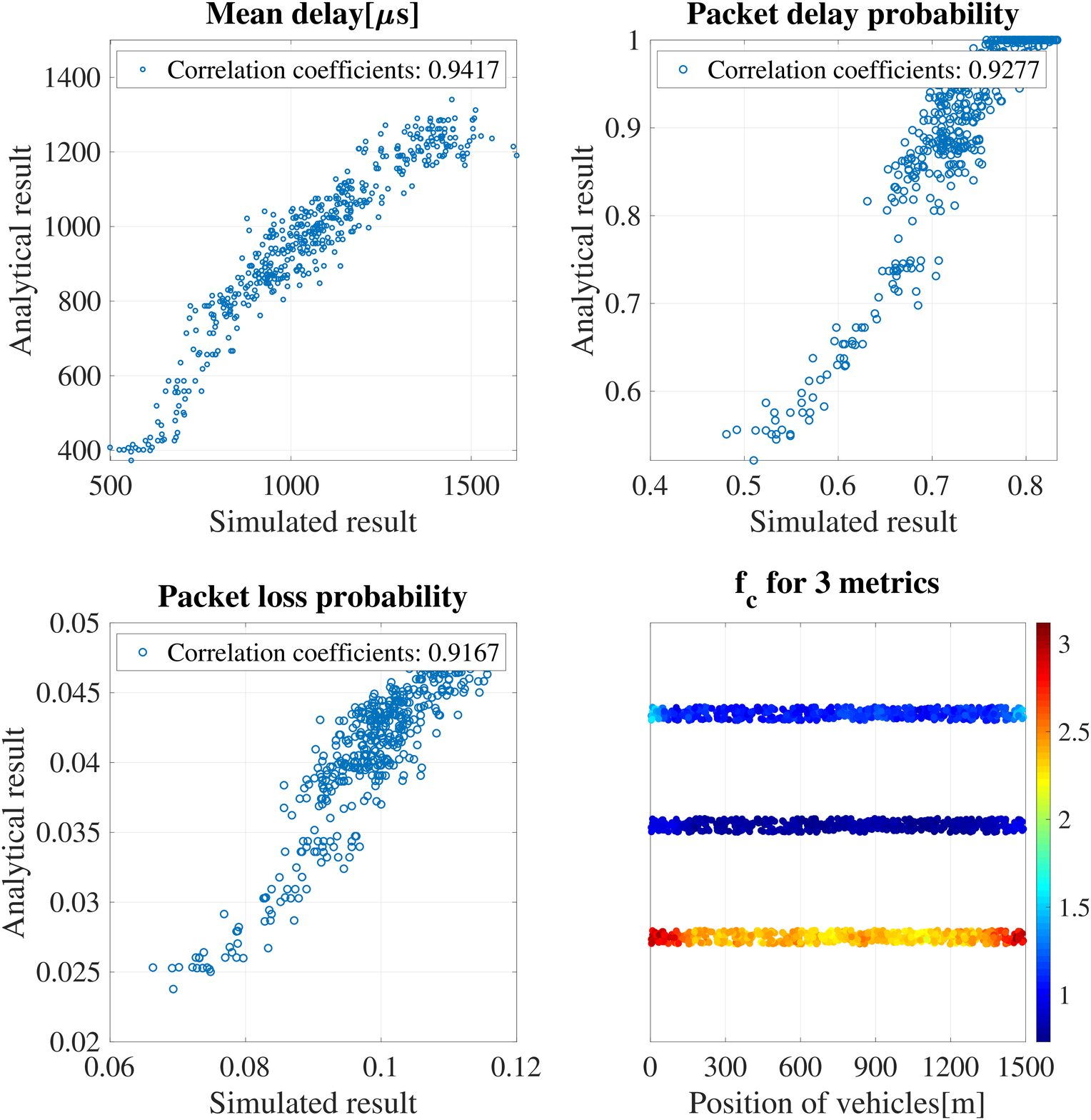}
      \caption{Correlation coefficients of 3 metrics are close to 1, which indicate a good feasibility of analytical estimation. To increase estimation accuracy, we introduce $\tilde{f_{c}}$}
      \label{fig:coefficients_3metrics}
   \end{figure}

Fig. \ref{fig:coefficients_3metrics} represents the correlation coefficients for each performance metric, which evaluate the strength of the association between simulated and analytical results. For an optimum fitting, the blue dots are supposed to be correctly distributed along the diagonal line, which denotes a correlation coefficient of 1. The correlation coefficients for the mean delay, packet delay probability, and packet loss probability are 0.9417, 0.9277 and 0.9167, which manifest a strong correlation and satisfying estimation ability of the analytical model.

To optimize the estimation performance of the proposed analytical model, the term correction factor ($f_c$) is introduced, 
\begin{equation}
\tilde{f_c} = \frac{\tilde{\Phi}_S}{\hat{\tilde{\Phi}}_A}\label{eq:CorrectionFactor}
\end{equation}
where $\tilde{\Phi}_S, \hat{\tilde{\Phi}}_A$ are the performance matrix from the simulation and the analytical model regarding the $t_{md}$, $P_{d}$, $P_{l}$, separately. 

Obviously, our goal can be demonstrated as Eq. \ref{eq:LossFunction}:

\begin{equation}
min(J) =\sum_{i}^{n=N} (\tilde{f_c} \cdot \hat{\tilde{\Phi}}_A - \tilde{\Phi}_S)^2  \label{eq:LossFunction}
\end{equation}
where N denotes the total number of vehicles.

The $ \tilde{\Phi}_S / \hat{\tilde{\Phi}}_A $ is shown in the bottom right sub-figure in Fig. \ref{fig:coefficients_3metrics}. The three curves from top to bottom indicate the $f_{l=c}$ for mean delay, packet delay probability and packet loss probability. The uniform color in the center area indicates that the naive analytical estimation method has stable performance and thus can be adjusted by multiplying appropriate correction factor $f_c$. Among 3 metrics, packet loss probability is dramatically underestimated and needs a larger $f_{c}$. This is because, in the LuT generation scenario, a reception is failed only due to multiple transmitter attempts to access the channel simultaneously, without consideration of hidden node. However, in the real-time simulation, the transmissions from the hidden nodes cause interference at the receiver. Consequently, the reception is more like to corrupt due to lower SINR.

The correction factor differs in the discontinuous edge of the scenario, where hidden node problem is not obvious. In this case, we introduce another correction matrix. Tab. \ref{tab:Correction factors} records the correction factor in the middle ($f_{c,c}$) and the correction factor at the edge ($f_{c,e}$), where the results are calculated based on \ref{eq:LossFunction}.

\begin{table}[!ht]
	\centering
	\caption{Correction factors}
	\begin{tabular}{l|cl}
	\hline \hline
	$f_c$
	    & center
		& edge \\ \hline
	Mean delay
		& 1.0857  
		& 1.3048\\
	Packet delay probability
		& 0.7516 
		& 0.9671\\
	Packet loss probability
		& 2.2617
		&  2.9121\\
	\hline \hline
	\end{tabular}
	\label{tab:Correction factors}
\end{table}

After using the correction factors, the analytical model outputs a very similar result to the simulation model. Furthermore, the LuT is portable to scenarios with different PHY parameters and path loss models, by re-calculating the transmission and sensing range size, since the contention mechanism due to CSMA/CA stays the same.

\section{Conclusion}

In this paper, we suggest that the IEEE 802.11p is a better solution for the first version of the fleet management of mobile working machines based on the analysis of the ad-hoc network and the cellular network. Moreover, we propose an analytical model to let mobile working machines have a real-time sense of the packet delay probability, mean delay and the probability of packet loss in the ad-hoc network. That is, the machine can estimate how probable its transmission can be delayed, how long its transmission can be delayed and how many packets can be lost in real-time. Thanks to V2X technology, mobile machines can work closer and be driven faster so that the productivity of the working site can be increased dramatically.   
However, our results also show the applicable conditions of IEEE 802.11p on mobile machines. As the nodes increase, the ad-hoc network may overload. Therefore, in our second version, we are going to publish a V2X solution that combines the IEEE 802.11p and 5G. In that version, machines use the analytical model proposed in this paper to decide when the 5G should be applied.
Due to the limit of the pages, we just introduce the core ideas and the results. The source code and generated LUT are available on \href{https://github.com/XiangYusheng/ConnectedMobileMachines.git}{Github}.\\

\addtolength{\textheight}{-12cm}   





\end{document}